\documentclass[aps,twocolumn,superscriptaddress]{revtex4}
\usepackage{graphics}
\begin{document}

\title{Predicting C-H/$\pi$ interactions with nonlocal density functional theory}

\author{Joe Hooper}
\email[Email address: ]{joseph.p.hooper@navy.mil}
\affiliation{Naval Surface Warfare Center, Indian Head, Maryland 20640}
\author{Valentino R. Cooper}
\author{T. Thonhauser}
\affiliation{Department of Physics and Astronomy, Rutgers University, Piscataway, New Jersey 08854}
\author{Nichols A. Romero}
\affiliation{U.S. Army Research Laboratory, Aberdeen Proving Ground, Maryland 21005}
\author{Frank Zerilli}
\affiliation{Naval Surface Warfare Center, Indian Head, Maryland 20640}
\author{David C. Langreth}
\affiliation{Department of Physics and Astronomy, Rutgers University, Piscataway, New Jersey 08854}

\begin{abstract}
We examine the performance of a recently developed nonlocal density functional in predicting a model noncovalent interaction, the weak bond between an aromatic $\pi$ system and an aliphatic C-H group. The new functional is a significant improvement over traditional density functionals, providing results which compare favorably to high-level quantum-chemistry techniques but at considerably lower computational cost. Interaction energies in several model C-H/$\pi$ systems are in generally good agreement with coupled-cluster calculations, though equilibrium distances are consistently overpredicted when using the revPBE functional for exchange. The new functional correctly predicts changes in energy upon addition of halogen substituents.
\end{abstract}
\maketitle

\section{Introduction}
Accurate treatment of van der Waals bonding represents one of the primary challenges for modern quantum chemical calculations. Such interactions are important in a great diversity of chemical and biological systems, and have been difficult to model in a highly accurate and computationally tractable manner. State-of-the-art techniques such as coupled cluster theory (CCSD(T)) generally treat such interactions very accurately, but are computationally demanding and often depend sensitively on the basis set used. Density functional theory (DFT) methods are attractive in terms of computational efficiency, but common exchange-correlation functionals used do not include effects such as dispersion interactions. For this reason, traditional DFT calculations frequently provide a poor description of van der Waals bonding in molecular systems.

Here we consider a recently developed van der Waals density functional (vdW-DF) which explicitly includes nonlocal electron correlation without the use of any empirical parameters \cite{Dion_vdw_PRL,Timo_Vxc}. This method has been successfully applied to a number of aromatic $\pi$ systems such as dimers of benzene \cite{Puzder_Benzene_JCP,Thonhauser_Benzene_JCP}, napthalene, anthracene, and pyrene \cite{Chakarova_PAHs_JCP}, as well as a variety of bulk systems such as crystalline polyethylene \cite{Kleis_Polyethylene}. The vdW-DF functional was strongly influenced by the physics included in its forerunner \cite{Rydberg_layered_PRB,Rydberg_layered_PRL,Langreth_IJQC}.

Of key importance is the ability of this nonlocal functional to treat a range of different noncovalent interactions. Along these lines we consider the attractive bonding between an aliphatic C-H group and an aromatic $\pi$ system, a configuration which has received considerable attention in recent years \cite{NishioBook,WeakHBond}. This so-called C-H/$\pi$ interaction is frequently referred to as a weak hydrogen bond, though recent studies have demonstrated that it arises primarily from dispersion, with electrostatic attraction playing a minor role \cite{Tsuzuki_Dispersion1,Tsuzuki_Dispersion2,Ringer}. The C-H/$\pi$ interaction is important in molecular packing \cite{CHpi_MolPac}, protein interactions \cite{CHpi_protein1,CHpi_protein2}, and in a variety of biological systems \cite{CHpi_biology1,CHpi_biology2}. The wide interest in this interaction, combined with the availability of high level CCSD(T) and MP2 calculations, make it a valuable test case for the new van der Waals functional.

In this paper we consider a number of model systems which allow careful study of the C-H/$\pi$ interaction. These include the methane-benzene system, trifluoro- and trichloromethane with benzene, and the methane-indole complex. All results are compared with high-level CCSD(T) and MP2 calculations. The interaction energies from vdW-DF are a significant improvement over gradient-corrected and hybrid DFT functionals, and compare favorably to CCSD(T) results extrapolated to the limit of a complete basis set. Equilibrium distances are generally overpredicted by vdW-DF . Our results suggest that this functional, which requires little time beyond a standard DFT calculation, holds great promise for modeling large organic and biological systems.

\begin{figure}
\includegraphics{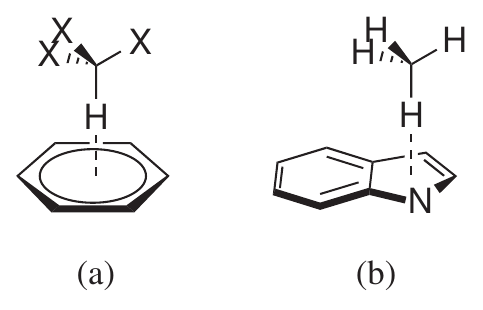}
\caption{The orientation of (a) CHX$_3$-benzene molecules, where X=H, F, or Cl, and (b) the methane-indole system. \label{Geometry}}
\end{figure}

\section{Nonlocal density functional}
Our approach to calculating the C-H/$\pi$ interaction uses a recent density functional which explicitly includes van der Waals correlations \cite{Dion_vdw_PRL,Timo_Vxc,Rydberg_layered_PRL,Langreth_IJQC,Rydberg_layered_PRB}. In this framework, which we will abbreviate as vdW-DF, the total energy functional is written as

\begin{equation}
E[\rho] = T_s[\rho] + V_{pp}[\rho] + E_H[\rho] + E_x[\rho] + E_c^L[\rho] + E_c^{NL}[\rho],
\end{equation} where $T_s$ is the independent-particle kinetic energy, $V_{pp}$ is the ionic pseudopotential, and $E_H$ is the Hartree energy. For the exchange-correlation functional, we separate the exchange energy $E_x$, the local correlation $E_c^L$, and the nonlocal component $E_c^{NL}$. Following previous work, we use the revPBE expression for exchange \cite{revPBE} for all calculations. $E_c^L$ is approximated here by the LDA correlation. The nonlocal term, which is key for capturing long-range correlations such as dispersion, can be expressed in a general form as 

\begin{equation}
E_c^{NL} = \frac{1}{2} \int \!\!\! \int d\mathbf{r}_1 d\mathbf{r}_2 \rho(\mathbf{r_1})\phi(\mathbf{r}_1,\mathbf{r}_2)\rho(\mathbf{r}_2).
\end{equation}

The kernel $\phi(\mathbf{r}_1,\mathbf{r}_2)$ depends only on the distance $|\mathbf{r}_1-\mathbf{r}_2|$ and the density $\rho$ near $\mathbf{r}_1$ and $\mathbf{r}_2$. Details regarding explicit evaluation of the kernel are given in Ref. \cite{Dion_vdw_PRL}. No empirical parameters are used in the nonlocal correlation. Though the revPBE exchange does not satisfy the Lieb-Oxford bound for any arbitrary density as does PBE, in practice it satifies the true, integrated Lieb-Oxford condition for typical systems \cite{revPBE,RPBE}. Thus our overall approach, using vdW-DF and revPBE exchange, may be considered a truly first-principles method for treating van der Waals systems.

In previous work $E_c^{NL}$ was computed as a post-processing step using the charge density from a normal DFT calculation \cite{Puzder_Benzene_JCP,Thonhauser_Benzene_JCP}. Explicit evaluation of the nonlocal potential $V_c^{NL}$ in a self-consistent way is possible through recently derived analytical expressions \cite{Timo_Vxc}. However, test calculations for the methane-benzene system indicated little difference in the interaction energies between these two approaches. Thus for all systems we continue to utilize the post-processing method, which requires very little time beyond a standard DFT calculation.

All vdW-DF calculations were done in a modified version of {\tt abinit} \cite{abinit}. Norm-conserving Troullier-Martins pseudopotentials were used, with a kinetic energy cutoff of 50 Rydberg. Calculations were performed in a large cubic cell with at least 18 \AA\ of space between adjacent periodic images. The nonlocal energy was calculated on a real-space grid identical to the one used for Fourier transforms in the planewave code. Calculations for hybrid DFT functionals were performed using {\tt CRYSTAL03} \cite{CRYSTAL03}.

Molecular geometries were optimized using the B3LYP hybrid functional with an aug-cc-pVTZ basis set; this optimization gave structures that were generally within 0.01 \AA\ of previous high-level calculations for methane-benzene and methane-indole \cite{Ringer}. These same geometries were also converged with respect to forces derived from the self-consistent implementation of vdW-DF, indicating that the nonlocal functional vanishes appropriately in intramolecular situations. In all cases the isolated molecule geometry was kept rigid during the two-molecule calculations.

\section{Results and discussion}

\subsection{Methane-benzene system}
The methane-benzene complex represents one of the simplest interacting C-H/$\pi$ systems. Following the results of previous experimental \cite{Schauer_Jet_JCP,JapaneseExp} and theoretical \cite{Tsuzuki_MethBenz,Ringer} work, the molecules are oriented such that a single C-H group on the methane points directly at the center of the benzene, as shown in Fig.~\ref{Geometry}. The separation is then defined as the distance between the central C atom in methane and the center of the benzene ring. The separation axis is thus coincident with the $C_6$ symmetry axis of benzene, and the dimer system overall is set in a $C_{3v}$ symmetry. Similar to the report by Ringer and coauthors \cite{Ringer}, the dimer energy is essentially invariant for rotations around the dimer separation axis, varying by at most $10^{-4}$ kcal/mol.

\begin{figure}
\includegraphics{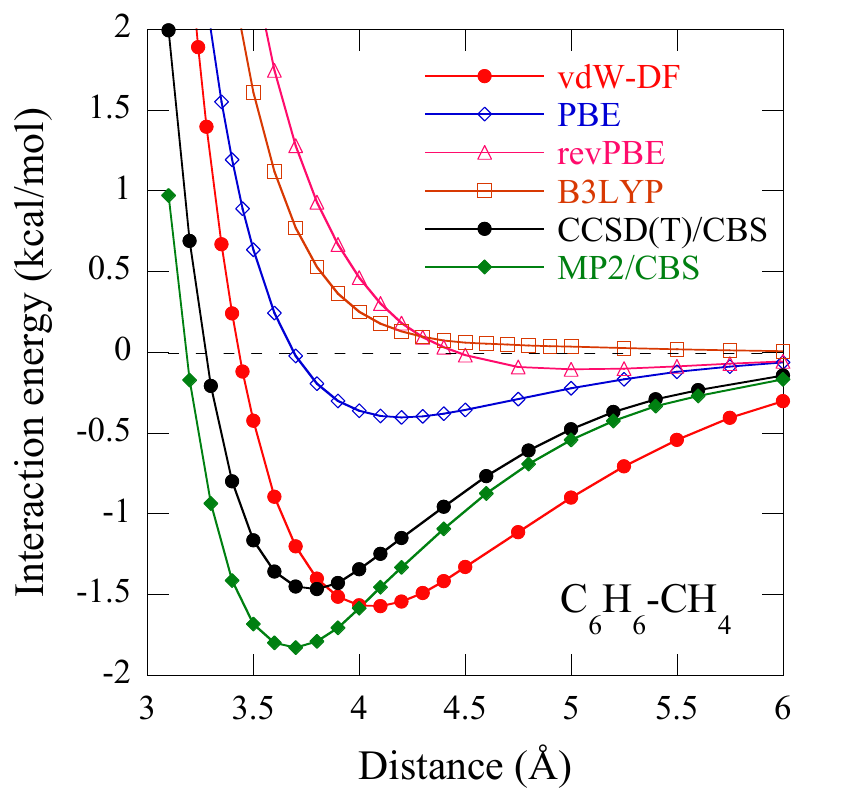}
\caption{The interaction energy of the methane-benzene dimer using vdW-DF. Comparison is made with the PBE \cite{PBE} and revPBE \cite{revPBE} GGA functionals, the B3LYP hybrid functional \cite{B3LYP}, as well as previously reported CCSD(T) and MP2 data extrapolated to the limit of a complete basis set (CBS) \cite{Ringer}. \label{MethaneBenzeneDimer}}
\end{figure}

In Fig.~\ref{MethaneBenzeneDimer} we present our results for the interaction energy of the methane-benzene dimer. The vdW-DF functional is a significant improvement over other standard DFT approaches, yielding an interaction energy comparable to the CCSD(T) results of Ringer \textit{et al} \cite{Ringer}. Their coupled-cluster interaction energy was extrapolated to the limit of a complete basis set (CBS) using Helgaker's method with large augmented correlation-consistent triple- and quadruple-$\zeta$ basis sets \cite{Ringer}. Shibasaki and coauthors also report similar theoretical CCSD(T) results \cite{JapaneseExp}. Despite its relative simplicity, vdW-DF captures the C-H/$\pi$ energy quite accurately as compared to these high level calculations. The equilibrium dimer separation predicted by vdW-DF is roughly 8\% larger than the CCSD(T) result, which appears to be a general trend of vdW-DF when using the revPBE functional for the exchange energy \cite{Puzder_Benzene_JCP,Thonhauser_Benzene_JCP,Timo_Vxc,Kleis_Polyethylene}.

Three results using traditional DFT methods are also shown in Fig.~\ref{MethaneBenzeneDimer}. The Perdew-Burke-Ernzerhof (PBE) nonempirical generalized gradient (GGA) functional produces a binding that is considerably weaker than that given by vdW-DF, though its predicted equilibrium distance is only slightly higher than the nonlocal functional (see Table \ref{IntEnergies} for numerical values). Zhang and Yang's revised PBE functional (revPBE), which is identical to PBE except for a semiempirical adjustment of one parameter in the exchange portion \cite{revPBE}, produces no binding. The popular hybrid functional B3LYP, which uses a semiempirical  mixing of exact exchange and GGA terms \cite{B3LYP}, also produces no binding for the system. Recent calculations by Shibasaki and coauthors show similar results for a range of GGA and hybrid functionals \cite{JapaneseExp}. Thus, as we would expect for a dispersion-dominated dimer, traditional DFT methods compare poorly to high level CCSD(T) and MP2 results.

Shibasaki and coworkers have recently measured the experimental interaction energy of gas phase methane-benzene clusters by mass analyzed threshold ionization spectroscopy \cite{JapaneseExp}. For comparison with the experimental value we must subtract the vibrational zero-point energy (ZPE) from our predicted C-H/$\pi$ interaction energy. The predicted change in ZPE for formation of the methane-benzene dimer is 0.296 kcal/mol using MP2 with a cc-pVTZ basis set and scaling factors to match the experimental and theoretical frequencies for the methane symmetric stretching mode \cite{JapaneseExp}. Subtracting this gives an interaction energy of $-$1.27 kcal/mol using vdW-DF and $-$1.15 kcal/mol using CCSD(T); both of these compare favorably to the experimental binding energy, which was determined to be between $-$1.03 and $-$1.13 kcal/mol \cite{Ringer,JapaneseExp}.

\subsection{Trifluoro- and trichloromethane}
\begin{figure}
\includegraphics{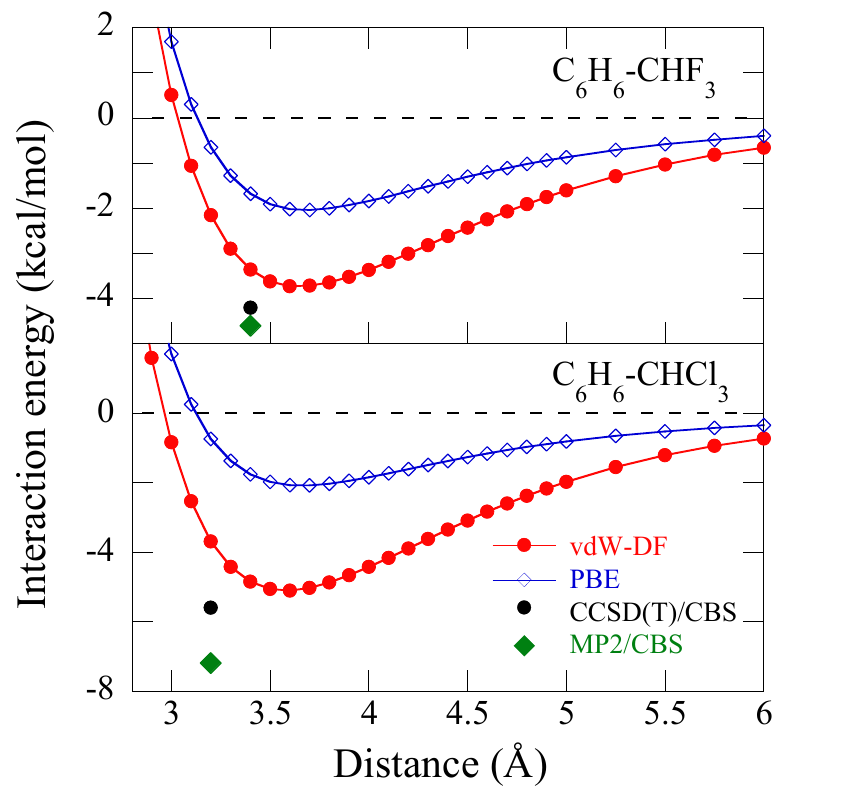}
\caption{The interaction energy of benzene with trifluromethane (top panel) and trichloromethane (bottom panel). Results are compared to previous CCSD(T) and MP2 calculations by Tsuzuki and coworkers extrapolated to the limit of a complete basis set (CBS) \cite{HalogenatedCHp}. \label{ChloroFluoroMethane}}
\end{figure}

We next consider the effect of halogen substituents on the aliphatic molecule. Some experimental measurements of the role of such substituent atoms on the C-H/$\pi$ interaction in solution have been reported, including evidence that benzene forms a stable dimer with trichloromethane \cite{Chloroform_Benzene1,Chloroform_Benzene2}. Here we consider replacement of three hydrogen atoms on the methane by chlorine or flourine, both of which increase the interaction energy of the complex. The results for these dimers of trifluoro- and trichloromethane with benzene are given in Fig.~\ref{ChloroFluoroMethane} and Table \ref{IntEnergies}.

In both systems the C-H bond points directly at the center of the benzene ring, and the remaining three H atoms on the methane are replaced. The geometries for CHF$_3$ and CHCl$_3$ were optimized independently, and the monomer structure was also used for dimer calculations. Trifluoro- and trichloromethane both have increased molecular polarizabilities compared to CH$_4$. Methane has a polarizability of roughly 17.9 a.u., compared to 19.0 a.u. for trifluoromethane and 57.6 a.u. for trichloromethane \cite{ExperimentalPolarize}. Since dispersion is the dominant component of the C-H/$\pi$ interaction, we would expect an increased interaction energy for the molecules with larger polarizabilities. This is indeed observed, as seen in Fig.~\ref{ChloroFluoroMethane}; the benzene-trichloromethane binding is much stronger than that seen in the methane system, and is comparable to the strength of the hydrogen bond in a dimer of water. We would also expect an increase in electrostatic attraction upon substitution of F or Cl, though previous results indicate this is a much smaller effect than the enhanced dispersion \cite{HalogenatedCHp}.

The performance of vdW-DF in these halogenated dimers is similar to that seen in the methane-benzene system. The nonlocal functional is a considerable improvement over the prototypical GGA functional PBE, providing interaction energies comparable to CCSD(T) results extrapolated to the limit of a complete basis set by Tsuzuki and coauthors \cite{HalogenatedCHp}. In these systems the vdW-DF interaction energy underestimates the previously reported CCSD(T) limit, in contrast to a slight overestimation for the methane-benzene and methane-indole complexes. The literature values for the latter two systems used slightly different basis sets as well as different methods for extracting the complete basis limit than were used for the halogenated dimers \cite{HalogenatedCHp,Ringer}. We note again the advantage of using a simple, straightforward DFT method such as vdW-DF.

The vdW-DF functional with revPBE exchange again overstimates the equilibrium separation of the dimers by approximately 6\% for trifluoromethane-benzene and 13\% for trichloromethane-benzene. MP2 and CCSD(T) calculations indicate that the latter has a smaller dimer separation, as would be expected from the increased dispersion component. vdW-DF does not reproduce this trend, predicting similar separations for both halogenated dimers.

\begin{table}
 \caption{Dimer separations (R$_{\rm D}$) and interaction energies E$_{\rm int}$ for all C-H/$\pi$ systems. Distances are in \AA\ and energies are in kcal/mol. vdW-DF results are given using the revPBE exchange functional. CCSD(T) and MP2 results are taken from Refs. \cite{Ringer} and \cite{HalogenatedCHp}. \label{IntEnergies}}
 \begin{ruledtabular}
 \begin{tabular}{lcc|cc|cc|cc}
 & \multicolumn{2}{c}{C$_6$H$_6$-} & \multicolumn{2}{c}{C$_6$H$_6$-} & \multicolumn{2}{c}{C$_6$H$_6$-} & \multicolumn{2}{c}{C$_8$H$_7$N-}\\
 & \multicolumn{2}{c}{CH$_4$} & \multicolumn{2}{c}{CHF$_3$} & \multicolumn{2}{c}{CHCl$_3$} & \multicolumn{2}{c}{CH$_4$}\\
 & R$_{\rm D}$ & E$_{\rm int}$ & R$_{\rm D}$ & E$_{\rm int}$ & R$_{\rm D}$ & E$_{\rm int}$ & R$_{\rm D}$ & E$_{\rm int}$\\
 \hline
vdW-DF & 4.1 & $-$1.57 & 3.6 & $-$3.73 & 3.6 & $-$5.11 & 4.0 & $-$1.81\\
PBE & 4.2 & $-$0.403 & 3.7 & $-$2.04 & 3.7 & $-$2.07 & 4.2 & $-$0.408 \\
CCSD(T) & 3.8 & $-$1.45 & 3.4 & $-$4.20 & 3.2 & $-$5.60 & 3.8 & $-$1.57 \\
MP2 & 3.8 & $-$1.79 & 3.4 & $-$4.60 & 3.2 & $-$7.20 & 3.7 & $-$1.96
 \end{tabular}
 \end{ruledtabular}
 \end{table}

\subsection{Methane-indole}

We next consider the interaction of methane with a larger aromatic system, the indole molecule. The methane is placed over the five-membered ring of indole, again with a single C-H bond pointing directly at the center of the ring (see Fig.~\ref{Geometry}). Geometries of the monomers were optimized separately and kept rigid during dimer calculations. In this configuration the interaction energy is again essentially invariant to rotations around the axis connecting the methane C and the center of the five-membered ring.

\begin{figure}
\includegraphics{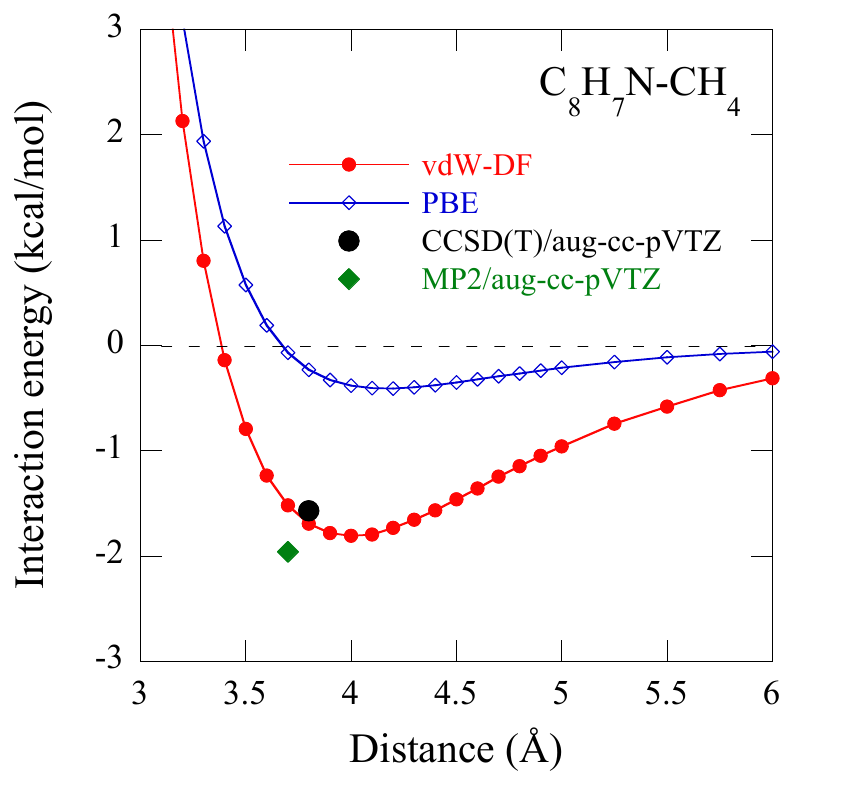}
\caption{The interaction energy of the methane-indole dimer. CCSD(T) and MP2 results are taken from Ref. \cite{Ringer}. \label{MethaneIndoleDimer}}
\end{figure}

Our results for the interaction energy of the methane-indole dimer are given in Fig.~\ref{MethaneIndoleDimer} and Table \ref{IntEnergies}. MP2 and CCSD(T) results are taken from the work of Ringer and coauthors, where an aug-cc-pVTZ basis set was used \cite{Ringer}. The magnitude of the interaction and the dimer separation are similar to the methane-benzene dimer, though vdW-DF predicts that methane-indole is more strongly bound, in agreement with CCSD(T) and MP2 results. An analysis by symmetry-adapted perturbation theory shows a similar distribution of components in the methane-benzene and methane-indole dimers, the most significant difference being a larger dispersion component (0.589 kcal/mol more negative) in the latter. As before, vdW-DF overpredicts the CCSD(T) dimer separation by approximately 5\%.

\subsection{Conclusion}

We have studied the performance of the van der Waals density functional in treating the weak bond between an aromatic $\pi$ system and an aliphatic C-H group. The properties of four prototypical C-H/$\pi$ systems were analyzed: methane-benzene, trifluoro- and trichloromethane with benzene, and methane-indole. 

The vdW-DF method is a substantial improvement over standard DFT functionals, providing in all cases an interaction energy that is comparable to high-level quantum chemistry calculations. In the case of the methane-benzene dimer, vdW-DF predicts a total interaction energy of $-$1.57 kcal/mol, compared to $-$1.454 kcal/mol from CCSD(T) and $-$1.79 kcal/mol from MP2 \cite{Ringer}. Traditional DFT functionals predict little or no binding for this dimer, and generally wavefunction-based methods require large basis sets to achieve values in this range \cite{JapaneseExp,Tsuzuki_MethBenz}.

The substitution of F and Cl for three of the methane hydrogens results in an increased binding due to larger dispersion and electrostatic interactions. The vdW-DF functional correctly reproduces the increased interaction energy in good agreement with the CCSD(T) trend, but does not match the reduction in equilibrium dimer separation. vdW-DF is also able to resolve the small binding energy difference between the methane-benzene and methane-indole dimers, but overpredicts the equilibrium separation in the latter system as well.

%Overall, equilibrium distances using the new functional are generally larger than CCSD(T) values by 5\%-15\% for these C-H/$\pi$ systems, which appears to be a consistent trend when using the revPBE exchange functional as part of the method \cite{Thonhauser_Benzene_JCP,Puzder_Benzene_JCP}. The predicted distances are only a slight improvement over the standard PBE GGA functional, though this behavior in part results from the choice of exchange functional. The approach to zero binding at larger dimer separations is also too gradual when using the revPBE exchange. Use of the PBE exchange, which differs from revPBE at intermediate and large reduced density gradients, produces a smaller equilibrium separation than CCSD(T) results but also an interaction energy that is far too large. RPBE exchange does not correct the deficiencies of the revPBE term, and also results an interaction energy larger in magnitude than the MP2 results. Further work is needed to determine the most suitable form of exchange to use with the nonlocal correlation in vdW-DF.

Overall, the performance of the new van der Waals density functional for these very weak bonds is encouraging. With further development, this computationally tractable method holds great promise for complex organic and biological systems where this manner of weak bonding plays a key role.

\subsection{Acknowledgements}
The authors would like to thank B. Rice, E. Byrd, and W. Mattson for useful discussions. J.~H. acknowledges an NRL/ASEE Postdoctoral Fellowship at the Naval Surface Warfare Center, Indian Head. N.~A.~R. was supported by an NRC Research Associateship Award at the U.S. Army Research Lab. Work at the DoD was supported by the Office of Naval Research. Work at Rutgers was supported in part by NSF Grant No. DMR-0456937.

\bibliography{Hooper_CHpi}

\end{document}